\documentstyle[rotate, psfig]{mn}

\title[VLBI polarisation observations of B1422+231]
{ 
Milliarcsec-scale polarisation observations of the
gravitational lens B1422+231
      }

\author[A.R.~Patnaik et~al.]
{A.R.~Patnaik$^1$, A.J.~Kemball$^2$, R.W.~Porcas$^1$, M.A.~Garrett$^3$\\
$^1$Max-Planck-Institut f{\"u}r Radioastronomie, 
Auf dem H{\"u}gel 69, D--53121 Bonn, FRG \\
$^2$National Radio Astronomy Observatory, PO Box 0, Socorro,
NM 87801, USA \\
$^3$Joint Institute for VLBI in Europe, Postbus 2, 7900~AA~Dwingeloo,
The Netherlands \\
}

\date{}

\begin{document}

\maketitle

\begin{abstract}
  
  We present polarisation observations of the gravitational lens
  system B1422+231 made at 8.4~GHz using the VLBA and the 100m
  telescope at Effelsberg. All four images of the quasar show
  structure on the milliarcsec scale.  The three bright images show
  tangential stretching as expected from lens models.  Some basic
  properties of gravitational lensing are exhibited by this system.
  The surface brightness of images A and B are the same and the parity
  reversal expected in image B is revealed, for the first time,
  by polarisation observations.  There is a large differential Faraday
  rotation between images A and B.
  
\end{abstract}

\begin{keywords}
gravitational lensing
\end{keywords}

\section{Introduction}

Radio polarisation observations can be used as a powerful tool in the
study of gravitational lenses. Such observations yield two useful
parameters - the distributions of polarised intensity (or degree of
polarisation) and of position angle (PA) of polarisation.  Both the
degree and PA of polarisation of a point in an object are unchanged in
its images by the action of a gravitational lens.  In gravitational
lens searches the equal degrees of polarisation of images can be used
to discriminate amongst lens candidates. However, the measured PAs of
polarisation need not be the same at any given frequency, since the
different ray paths of the images can undergo different amounts of
Faraday rotation.  After correcting for the rotation measure (RM), the
intrinsic PA (PA at zero wavelength) must be the same for the lensed
images.

Any difference in RM between lensed images can give clues to the
nature of the lensing galaxy. For a gas-rich lens or a spiral galaxy
lens, the RMs of each lensed image (and also their difference) are
expected to be large compared to those from an elliptical galaxy lens.
The difference in RMs is assumed to be caused by the interstellar
medium of the lens.

The polarisation angle is not altered by the gravitational deflection
even though the total intensity (Stokes parameter I) distribution can
be distorted (Dyer \& Shaver 1992).  Milliarcsec-scale observations of
the gravitational lens B0218+357 have been used to demonstrate this
feature of the gravitational potential (Patnaik, Porcas \& Browne
1995). This property can help identify corresponding features in
distorted lensed images, even when this is unclear from total
intensity maps, since their degrees of polarisation must be equal, and
their PAs of polarisation must be parallel (after correction for any
differential RM).

The gravitational lens system B1422+231 was discovered by Patnaik et
al. (1992); it is a 4-image system with maximum image separation of
1.3~arcsec (Fig. 1). The background radio source is associated with a
15.5~mag quasar at a redshift of 3.62. The lensing galaxy has a
redshift of 0.338 (Kundi{\'c} et~al. 1997, Tonry 1998). The lensed
images have similar spectra at radio as well as optical wavelengths.
The three bright images, A, B and C, have similar radio polarisation
properties; image D is too weak at radio wavelengths for its
polarisation properties to be determined reliably. The source has been
observed in IR and optical bands (Lawrence et al. 1992; Remy et al.
1993; Yee \& Ellingson 1994, Bechtold \& Yee 1995; Akujor et al. 1996;
Yee \& Bechtold 1996; Impey et al. 1996) and models of the lensed
system have been proposed by several authors (Hogg \& Blandford 1994;
Narasimha \& Patnaik 1994, Kormann, Schneider \& Bartlemann 1994; Mao
\& Schneider 1998).

In this paper we describe high resolution radio polarisation
observations of the gravitational lens B1422+231, made at 8.4~GHz
using the VLBA together with the Effelsberg radio telescope.  The
observations and data reductions are described in Section 2; the
results and their implications are discussed in Section 3.

\begin{figure}
\raisebox{-3.2cm}
{\begin{minipage}{0.3cm}
\mbox{}
\parbox{0.3cm}{}
\end{minipage}}
\begin{minipage}{7cm}
\mbox{}
\rotate[r]{\psfig{figure=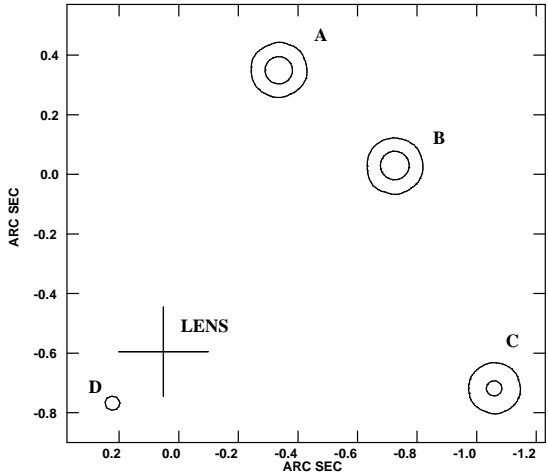,width=6.5cm,rheight=6.5cm,rwidth=6.5cm}}
\centering
\end{minipage}
\caption{Location of lensed images (in circles) and the lensing galaxy
  (marked + ) in B1422+231.  
}
\end{figure}

\section{Observations and Analysis}

We observed B1422+231 on 1997 June 11/12 at 8.4~GHz, using all 10
antennas of the VLBA and the 100m telescope at Effelsberg.  We
recorded eight 8~MHz channels using a dual polarisation set$-$up,
giving a total
bandwidth of 32~MHz in each polarisation; we used 1-bit sampling of
the signals.  The data were correlated at a single field centre using
the VLBA correlator.  The sampling in time was 1~sec and in frequency
0.5~MHz.  Since the image separations are large (the largest is
1.3~arcsec) compared to the synthesised beam (about 1~milliarcsec), it
is important to preserve short time and frequency sampling in the
analysis, as this avoids smearing of the visibility function, and
hence distortion of the images near the edge of the field.

We observed in cycles consisting of 6 mins. on the calibration source
1308+326 and 16 mins. on B1422+231, resulting in a total integration
time on B1422+231 of 410~minutes.  In between we observed the
calibrator sources 1226+023 (3C~273), 1749+096 and 1823+568 once each
for 8~min.

\begin{table*}
\begin{center}
\begin{tabular}{lrrlrrrr}

Image    & Total flux density & Polarised intensity & Deconvolved size  &  $\Delta$RA$_1$ &
         $\Delta$Dec$_1$ & $\Delta$RA$_2$ & $\Delta$Dec$_2$ \\
         & in mJy     & in mJy  & in mas$\times$mas, PA  & in mas & in mas &
         in mas & in mas \\
         &             &  &                     &   & & &          \\
A        & 152$\pm$2 & 3.7   & 2.11$\times$0.37, 53$^{\circ}$  & 389.25 &
         319.98 & 389.29 & 320.00\\
B        & 164$\pm$2  & 2.9  & 2.52$\times$0.28, 43$^{\circ}$  & 0  & 0 & 0
         & 0\\
C        & 81$\pm$1   & 0.95  & 1.57$\times$0.45, 16$^{\circ}$  & $-$333.88
         & $-$747.71 & $-$333.86 & $-$747.71 \\
D        & 5$\pm$0.5 &    & 0.89$\times$0.59, 123$^{\circ}$ & 950.65 &
         $-$802.15 & 950.67  & $-$802.12 \\
         &            &  &                    &  & & & \\
             
\end{tabular}
\end{center}
\caption{
  Total and polarisation flux densities and deconvolved sizes of the
  lensed images of B1422+231. Flux densities are given in mJy, and the
  deconvolved sizes are given in terms of major and minor axes (in
  mas) of the fitted elliptical gaussian and the PA (in degrees) of
  the major axis. Polarisation flux is the sum over the image area.
  The image separations are given in mas with respect to image B: the
  columns (5 and 6) with subscript 1 have been determined by JMFIT
  which fits an elliptical gaussian to the entire image and the
  columns (7 and 8) with subscript 2 have been determined by MAXFIT
  which fits a quadratic to the peak in the image.
  }
\end{table*}

The data were analysed using the NRAO software package {\sc AIPS}.
The data were corrected for the parallactic angle variations of the
telescopes, and amplitude calibrations determined from the measured
system temperatures and gains were applied. The phase slopes across
the 8~MHz bands were aligned using the pulse-cal information from each
telescope. Standard fringe-fitting was then performed on the
calibration sources.  After careful editing, we used the data on
1308+326 and 1823+568 to determine the bandpass function.

We used the following procedures to determine the instrumental
polarisation terms and correct for them.  After applying the total
intensity calibration table, we fringe-fit the cross-hand (RL, LR)
data of 1308+326 on the single baseline Los Alamos (our reference
antenna) to Pie Town.  In this way one determines the residual LHC-RHC
delay offset for the reference antenna. The solutions were smoothed,
copied to the calibration table and applied to the main data.

Since the sources used for polarisation calibration usually have
structures at milliarcsec-scales, a model of the source is required in
order to determine the instrumental polarisation leakage terms.
1308+326 was mapped to determine its structure and the model was used
in the AIPS task LPCAL.  1308+326 had a peak polarisation of 3.6
percent, and typical instrumental polarisation terms were 1 to 2 percent
for the VLBA telescopes and 4 percent for Effelsberg.

\begin{figure}
\raisebox{-3.2cm}
{\begin{minipage}{0.3cm}
\mbox{}
\parbox{0.3cm}{}
\end{minipage}}
\begin{minipage}{7cm}
\mbox{}
\psfig{figure=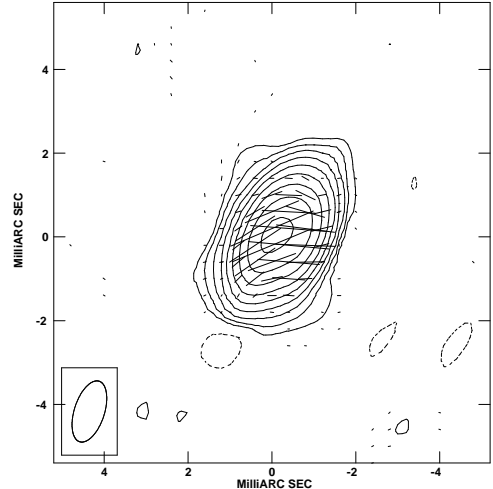,width=6.5cm,rheight=6.5cm,rwidth=6.5cm}
\centering
\end{minipage}
\caption{ 8.4~GHz map of 1308+326 with polarisation overlaid as vectors
  proportional to the polarised intensity.
  The contour levels are $-$2, $-$1, 1, 2, 4, 8, 16, 32, 64, 128 $\times$
  the contour interval which is chosen as the 3$\sigma$ noise in the
  map. The contour interval is 4.83~mJy~beam$^{-1}$ and the peak in
  the map is 1.514~Jy~beam$^{-1}$. The integrated flux density is
  2.648~Jy. The convolving beam of 1.52$\times$0.72~mas in position
  angle $-$17.8$^{\circ}$ is drawn at the lower left-hand corner of the map.}
\end{figure}

We have not been able determine the absolute angle of polarisation. In
Fig. 2 we show the map of 1308+326 with electric vectors overlaid on
the total intensity contours; the length of the vector is proportional
to polarised intensity.  The PA of polarisation changes within the
source, but from the integrated flux density of Q and U Stokes
parameters we find that the PA of 1308+326 was 11$^{\circ}$. We do not
have polarisation PA measurements of 1308+326 sufficiently close to
our observing epoch to calibrate the absolute angle of polarisation.

After performing the above polarisation calibration procedures, the
data for B1422+231 were analysed.  Fringe-fitting sources with widely
separated components, such as B1422+231, can prove difficult,
especially if no component consistently dominates on all baselines
(see Porcas, 1994). This is often the case for gravitational lenses,
since the equal surface brightness property of lensed images tends to
result in roughly equal contributions from them for baselines on which
they are resolved.  The AIPS fringe-fitting task FRING can utilise a
source model, and this feature can be used to guide the process.

We used the following procedure for the B1422+231 data.  First the
data are fringe-fitted using a point source model, and after applying
the solutions, a map is made using the AIPS tasks IMAGR and CALIB.
Since the lensed images are widely separated, we map each image in a
separate sub-field.  Even though this map is inaccurate, the images
can be identified at their correct locations.  A second fringe-fit is
then made, using a model comprising components from each of the
images, and a new map made. This process is repeated a few times to
converge on a consistent fringe-fit solution.

For the final map of the 4 image subfields, several cycles of phase
self-calibration were performed. Polarisation maps were made using
usual procedures.

\section{Results and Discussion}

Our maps of the 4 images of B1422+231, made using uniform weighting,
are shown in Figures 3 and 4.  The two strongest images, A and B, have
highly elongated structures, confirming the basic image shapes derived
from VLBA 15 GHz observations by Patnaik and Porcas (1998).  There is
only a single peak in the total intensity distributions of the images,
so we fit a single elliptical Gaussian function to each image using
task JMFIT. The 3 strongest images are many resolution elements in
length, and the fits are not particularly good, but they do
parameterize some basic image properties.  The total flux densities
(column 2), polarised intensity (column 3), deconvolved sizes (column
4) and relative positions (columns 5 and 6) are listed in Table 1. As
the strongest images are highly elongated, we have also made estimates
of the (local) positions of the central peaks using MAXFIT (columns 7
and 8 in Table~1).  However, there is no obvious compact feature
within the smooth image structures, and the peak positions are thus
not well defined, especially in the direction of elongation of the
images.  We therefore estimate an uncertainty in defining the image
positions of about 1/20th of the image size in the corresponding
direction.

Kochanek, Kolatt \& Bartelmann (1996) have suggested that repeated
measurements with $\sim$10$\mu$arcsec precision of the separation
between highly magnified image pairs such as A and B, would yield a
measurement of the proper motion of the lens with respect to the
observer.  Movement of the lens can result in a detectable image
motion due to magnification along the tangential direction.  The
B1422+231 system may prove difficult to use for such proper motion
studies, since the elongation of the relatively featureless images
results in an increased position uncertainty in the same direction.
This may also apply to the bright `merging image' pairs of other
4-image gravitational lens systems if they do not contain highly
compact features.

None of the images exhibits an obvious asymmetric structure of the
canonical "core-jet" type in its total intensity distribution.  Whilst
such morphologies occur frequently amongst radio-loud quasars, it
should be noted that the radio spectrum of B1422+231 is peaked around
5~GHz, and the relatively steep spectrum at higher frequencies
(Patnaik et al. 1992, Patnaik et al. 1999, in preparation) does not
show any evidence for "core dominance".  In any case, the
magnification and distortion of the background source structure by the lens
must also be taken into account.  Images A and B are highly magnified,
and image C is magnified by a factor of a few in most lens models.
Without the presence of the lens, B1422+231 would most likely be a
19~mag quasar with a radio flux density at 8.4~GHz of only ca 20~mJy -
almost a "radio quiet" quasar.

We obtain a number of important results from the total intensity
distribution.  We have investigated the surface brightness of the 2
strongest images, A and B, by measuring the flux in each image
contained above the contour level of 1~mJy~beam$^{-1}$. The flux ratio
of 0.94$\pm$0.02 is essentially the same as the area ratio of
0.95$\pm$0.05, thus demonstrating that the surface brightness is the
same in these two lensed images.  This accords with the property
that gravitational lensing does not change the surface brightness of
an image.

A second result is that the PA of elongation of the structures of A, B
and C are tangential with respect to the lens, which is believed to be
located close to image D (Fig. 1). This is expected
from lens models where the background source lies close to the diamond
caustic. Even though image D is weak, it is interesting to note that
its elongation appears `radial' with respect to the lens.

\begin{figure*}
\raisebox{-3.2cm}
{\begin{minipage}{0.3cm}
\mbox{}
\parbox{0.3cm}{}
\end{minipage}}
\begin{minipage}{8cm}
\mbox{}
\psfig{figure=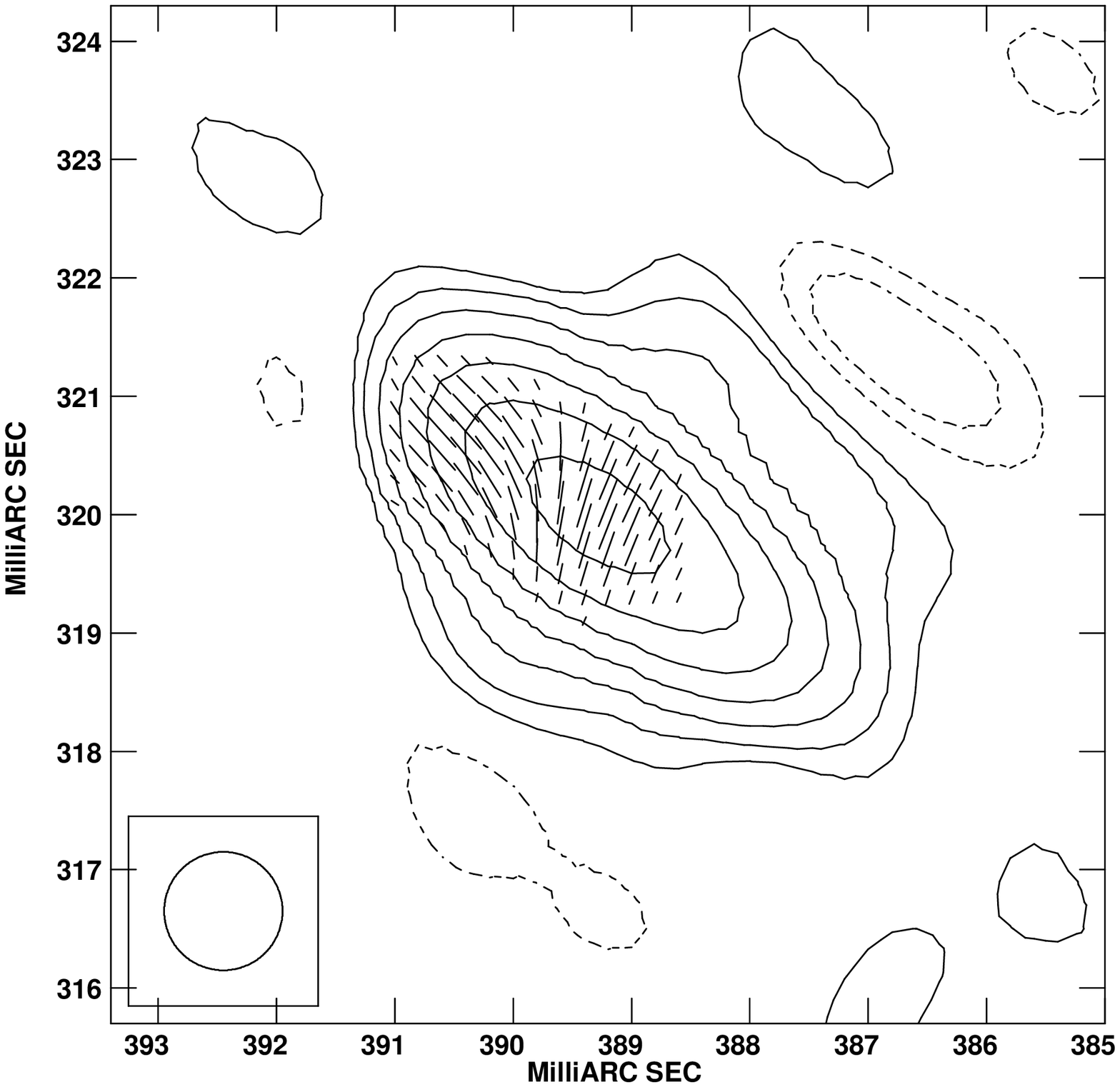,width=8cm,rheight=8cm,rwidth=8cm}
\centering
\end{minipage}
\hspace{0.5cm}
\raisebox{-3.2cm}
{\begin{minipage}{0.3cm}
\mbox{}
\parbox{0.3cm}{}
\end{minipage}}
\begin{minipage}{8cm}
\mbox{}
\psfig{figure=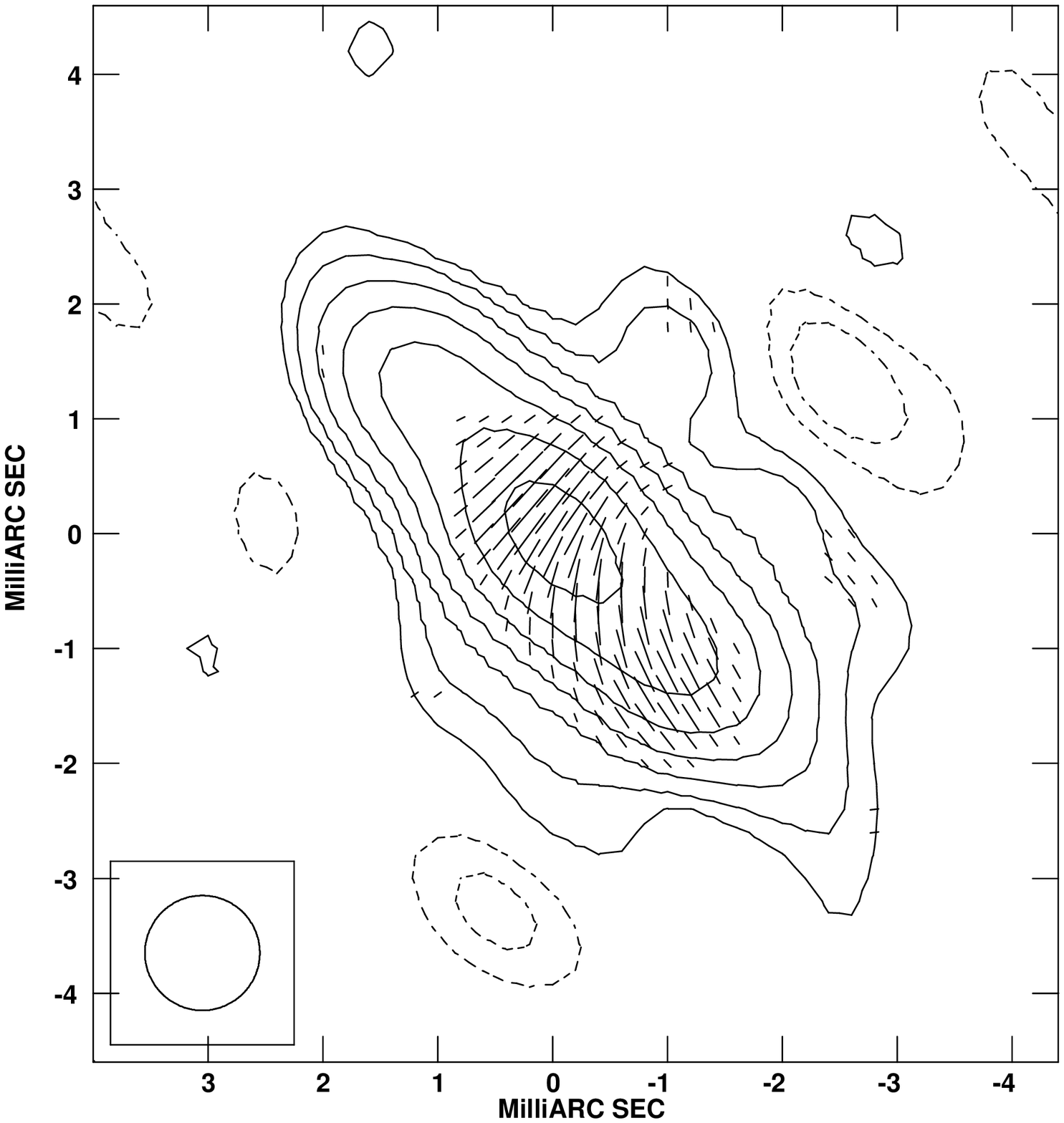,width=8cm,rwidth=8cm,rheight=8cm}
\centering
\end{minipage}
\vspace{0.5cm}
\caption{ 8.4~GHz map of B1422+231 A (left) and B (right). The contour
  levels for all the images are $-$2, $-$1, 1, 2, 4, 8, 16, 32, 64,
  128 $\times$ the contour interval which is chosen as the 3$\sigma$
  noise in the map.  For image A, contour interval is
  0.72~mJy~beam$^{-1}$ and peak flux density is 62.9~mJy~beam$^{-1}$
  and for image B contour interval is 0.72~mJy~beam$^{-1}$ and the
  peak is 64.6~mJy~beam$^{-1}$. The convolving beam of 1~mas circular
  gaussian is drawn at the lower left-hand corner in each map.
  Polarisation is plotted as electric vectors proportional to the polarised
  intensity.  }
\vspace{0.5cm}
\raisebox{-3.2cm}
{\begin{minipage}{0.3cm}
\mbox{}
\parbox{0.3cm}{}
\end{minipage}}
\begin{minipage}{8cm}
\mbox{}
\psfig{figure=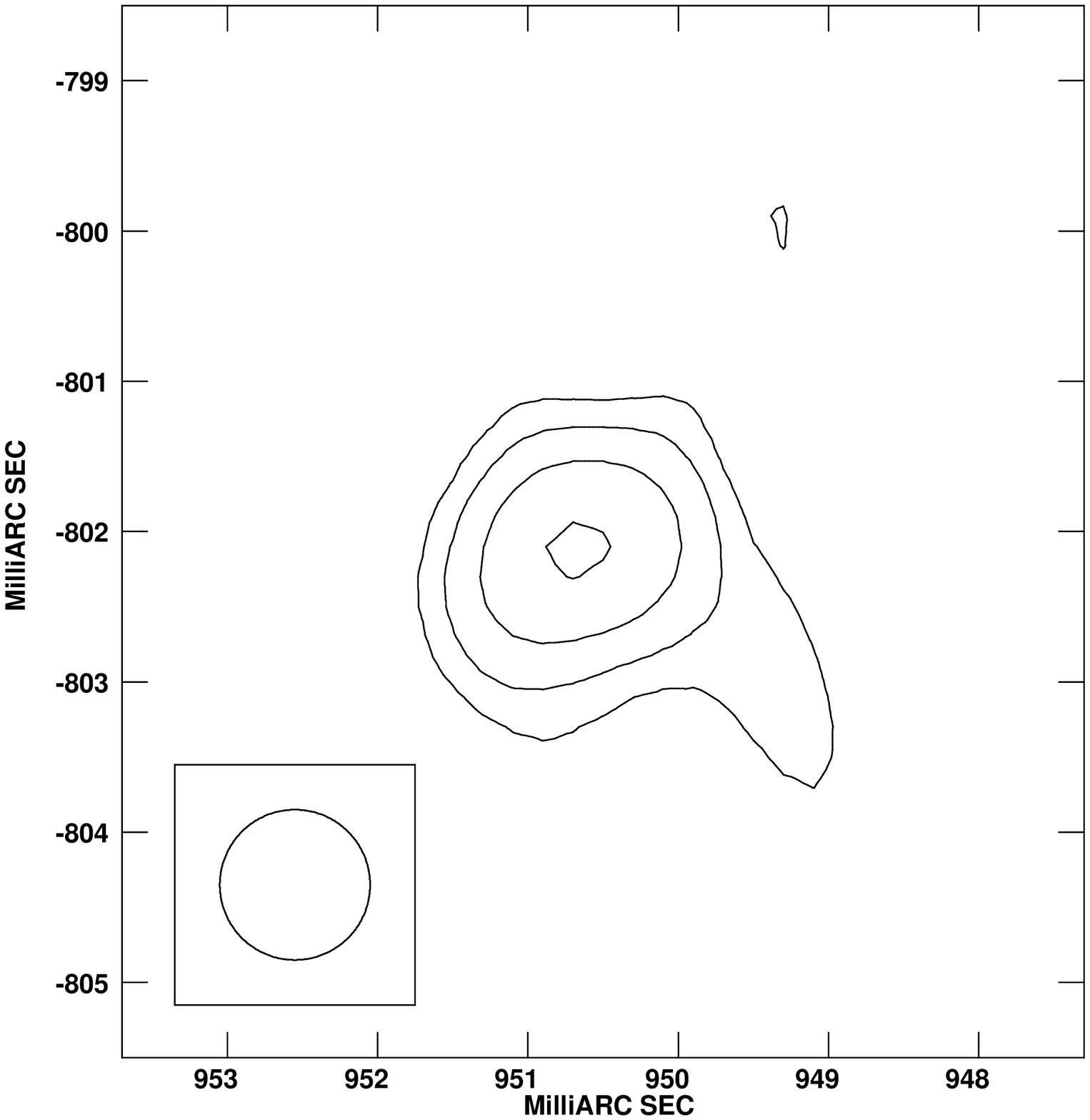,width=8cm,rheight=8cm,rwidth=8cm}
\centering
\end{minipage}
\hspace{0.5cm}
\raisebox{-3.2cm}
{\begin{minipage}{0.3cm}
\mbox{}
\parbox{0.3cm}{}
\end{minipage}}
\begin{minipage}{8cm}
\mbox{}
\psfig{figure=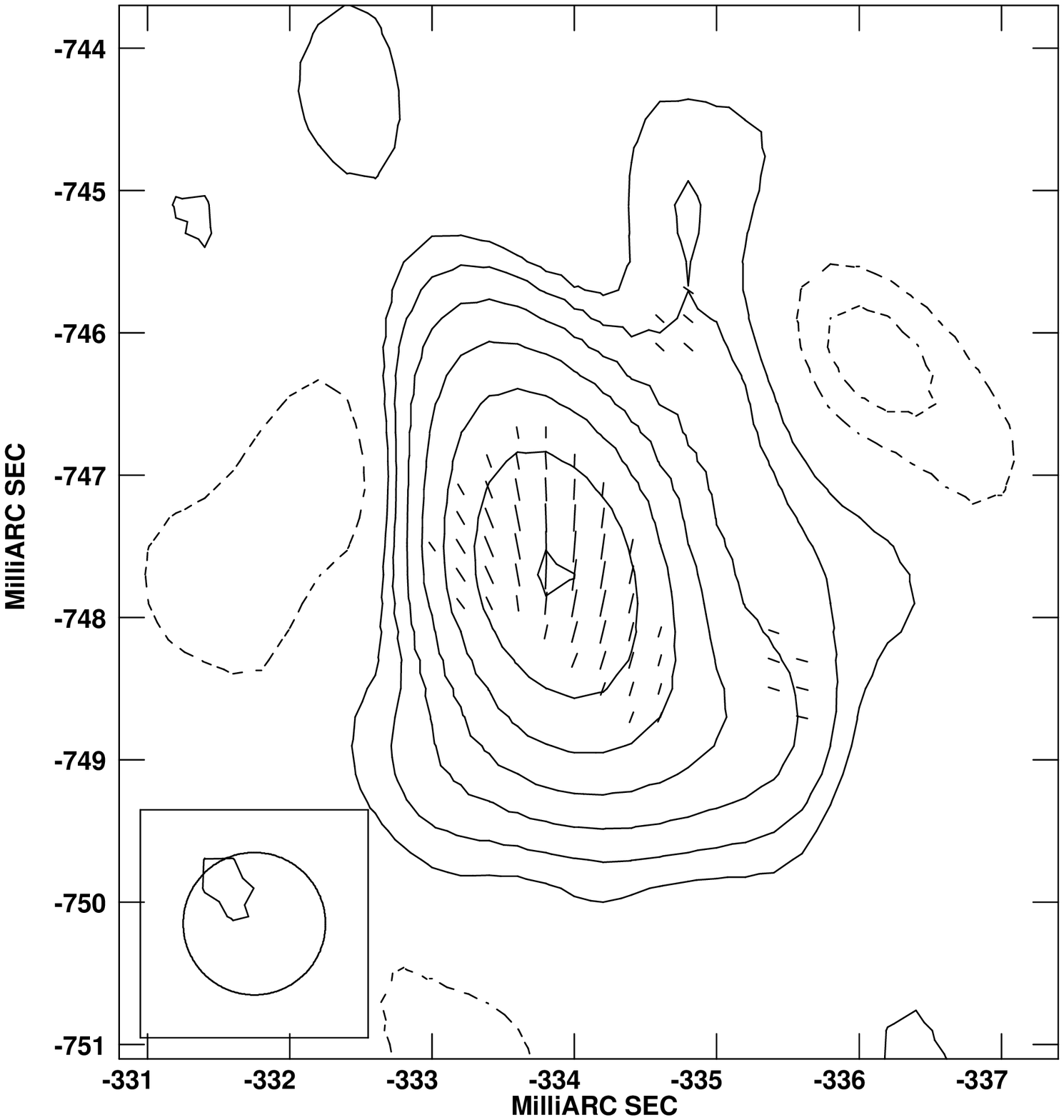,width=8cm,rwidth=8cm,rheight=8cm}
\centering
\end{minipage}
\vspace{0.5cm}
\caption{ 8.4~GHz map of B1422+231 C (right) and D (left). For image C,
  the contour interval is 0.62~mJy~beam$^{-1}$ and the peak flux
  density is 41.8~mJy~beam$^{-1}$. For image D, the contour interval
  is 0.36~mJy~beam$^{-1}$ and the peak flux density is
  3.1~mJy~beam$^{-1}$. 
  The convolving beam of 1~mas circular gaussian
  is drawn at the lower left-hand corner in each map. Polarisation is
  plotted as electric vectors proportional to the polarised intensity.
  }
\end{figure*}

The polarisation distribution in the images is more
remarkable.  Although we do not have enough sensitivity to detect
polarised emission from the weak D image, the polarisation
distribution in the other 3 images is clearly non-uniform, and the PA
of polarisation changes over the images.  In particular, there is a
"reflection symmetry" between A and B in the polarisation distribution
along the image major axes.  The SW of A and NE of B are regions of
little or no polarisation.  Progressing from these points along the
image axes, the degree of polarisation rises to $\sim$2.5\% at the
central peaks, with polarisation angles of $-$23$^{\circ}$ in A and
$-$43$^{\circ}$ in B.  Progressing further, the PAs rotate to final
values of 40$^{\circ}$ in the NE of A, and 22$^{\circ}$ in the SW of
B.
 
Since the polarisation properties of the source are not changed by
gravitational lensing, we may use such matching features in these
images to identify corresponding regions.  Thus the run of
polarisation emission in opposite directions in A and B reveals the
opposite orientation of these images in the tangential direction.
This is entirely expected, since these two bright images must have
opposite parity.  Indeed, the slight Southerly offset of the
polarisation peaks (in the NE of A and SW of B) from the axis of the
total intensity distributions establishes their opposite parity
directly.  Even though the polarisation of C is weak, the change of
polarisation within the image matches closely with that of image A;
this is expected as A and C should have the same parity.

From the measurements made on the images, noted above, it is clear
that there is a systematic difference of 20$^{\circ}$ between the
polarisation PAs of corresponding features in A and B.  Since the
gravitational action of the lens does not change the polarisation
angle of regions of the object in either image, we attribute this
observed difference to Faraday rotation along one or both image paths;
the difference in RM amounts to 280$\pm$20~rad~m$^{-2}$. Images A and
B are considered to be "merging images" and hence their ray paths,
separated by 0.5~arcsec and located on the same side of the lens, are
expected to traverse similar environments in the lens. Moreover, these
images are located some distance, (ca 1~arcsec) from the lens centre.
The differential RM is assumed to be caused by the magneto-ionic
medium of the lens. Such high RMs are generally found in gas-rich
environments; thus it is rather surprising that this lens, thought to
be an elliptical galaxy, can give rise to this large differential
Faraday rotation.

VLBI measurements of lensed image structures can potentially provide
more constraints for lens modelling than image relative flux densities
and positions. These are conveniently presented by the relative
magnification matrix between image pairs. Ideally there would be at
least 3 non-collinear points in the source, recognisable in each
image; the matrix could then be determined by considering the
transformation of 2 non-parallel vectors.

In B1422+231 there is only a single peak in the total intensity
distribution, but there are separate and distinct peaks in the
polarised flux distributions of A and B, which we recognise as
corresponding points from their polarisation properties.  We measured
the positions of these peaks in the maps of the A and B images and
attempted to determine an A/B relative magnification matrix.  However,
we were unable to get values for the matrix elements consistent with
the relative flux densities of the images.  The elements derived by
this method are in any case ill-defined, because the three peaks used
are almost collinear.

We have used the separations between the polarisation peaks within the
A and B images to test the matrices given by Hogg \& Blandford (1992).
Our measured separation in B is 1.478~mas in PA $-$145.2$^{\circ}$ .
Using their two matrices, we find that this transforms to a predicted
separation in image A of 1.607~mas in PA 27.9$^{\circ}$, and 1.680~mas
in PA 22.8$^{\circ}$ respectively. The measured separation is 1.0~mas
in PA 62.6$^{\circ}$.  It is perhaps not surprising that these models
fail in their predictions, as they are based on an measured optical
flux ratio between A and B of 0.77.  Transformation matrices are not
explicitly given for the other models, and so we have not tested them.
However, all the models predict tangential stretching of the three
bright images.

\section{Conclusion}

We have presented polarisation observations of the gravitational lens
system B1422+231 made using the VLBA and Effelsberg at 8.4~GHz. Our
1~mas resolution maps of the images reveal the parity reversal of
image B through the distribution of polarised emission.  We show that
the surface brightness is the same in A and B, as expected from
preservation of the source surface brightness.  We find that the differential
Faraday rotation between A and B is rather large, considering that the
lens is an elliptical galaxy. It is difficult to derive a
relative magnification matrix from the total intensity and
polarisation distributions of the A and B images. Published matrices,
however, do not successfully predict the structural relationships
between A and B.

\subsection*{ACKNOWLEDGMENTS}

We thank D.~Narasimha for helpful discussion and J.~Schmid-Burgk for
critical comments.  The National Radio Astronomy Observatory is a
facility of the National Science Foundation operated under cooperative
agreement by Associated Universities, Inc.


\def\beginrefer{\section*{References}%
\begin{quotation}\mbox{}\par}
\def\refer#1\par{{\setlength{\parindent}{-\leftmargin}\indent#1\par}}
\def\endrefer{\end{quotation}}

\beginrefer

\refer Akujor, C.E., Patnaik, A.R., Smoker, J.V., Garrington, S.T.,
1996, in Kochanek, C.S, Hewitt, J.N., eds,  Proceedings of the 173rd
Symposium of the IAU `Astrophysical applications of gravitational
lensing', Kluwer Academic Publishers, Dordrecht, p335 \\
\refer Bechtold, J., Yee, H.K.C.,  1995, AJ, 110, 1984\\
\refer Dyer, C.C., Shaver, E.G.,  1992, ApJ, 390,  L5 \\
\refer Hogg, D.W., Blandford, R.D.,  1994, MNRAS, 268,  889\\
\refer Impey, C.D., Foltz, C.B., Petry, C.E., Browne, I.W.A., Patnaik, 
A.R., 1996, ApJ, 462, L53\\
\refer Kochanek, C.S., Kolatt, T.S., Bartelmann, M., 1996, ApJ, 473, 610\\
\refer Kormann, R., Schneider, P., Bartelmann, M., 1994, AA, 286,  357\\
\refer Kundi{\'c}, T., Hogg, D.W., Blandford, R.D., Cohen, J.G., Lubin,
 L.M., Larkin, J.E.,  1997, AJ, 114, 2276\\
\refer Lawrence, C.R., Neugebauer, G., Weir, N., Matthews, K., Patnaik,
 A.R., 1992, MNRAS, 259, 5P\\
\refer Mao, S., Schneider, P., 1998, MNRAS, 295, 587 \\
\refer Narasimha, D., Patnaik, A.R., 1994, in Surdej, J.,
Fraipont-Caro, D, Grosset, E., Refsdal, S., Remy, M., eds, Proc. 31st
Li{\`e}ge International Astrophysical Colloq.,  Gravitational
Lenses in the Universe, Universit{\'e} de Li{\`e}ge, Belgique, p.295\\
\refer Patnaik, A.R., Browne, I.W.A., Walsh, D., Chaffee, F.H., Foltz, 
C.B.,  1992, MNRAS, 259, 1P \\
\refer Patnaik, A.R., Porcas, R.W., 1998, in Zensus, J.A., Taylor,
G.B., Wrobel, J.M., eds, Radio Emission from Galactic
and Extragalactic Compact Sources, ASP Conference Series, Volume 144,
IAU Colloquium 164, p319 \\
\refer Patnaik, A.R., Porcas, R.W., Browne, I.W.A., 1995, MNRAS, 274, L5\\
\refer Porcas, R.W., 1994, in Zensus, J.A., Kellermann, K.I., eds,
`Compact Extragalactic Radio Sources', NRAO, p125 \\
\refer Remy, M., Surdej, J., Smette, A., Claeskens, J.-F.,  1993, AA,
278, L19 \\
\refer Tonry, J.L., 1998, AJ, 115,  1 \\
\refer Yee, H.K.C., Bechtold. J., 1996, AJ, 111, 1007 \\
\refer Yee, H.K.C., Ellingson, E.,  1994, AJ, 107, 28 \\

\endrefer


\bsp

\end{document}